\documentclass{aa}
\usepackage{graphicx}
\usepackage{txfonts}
\begin{document}
   \title{The assessment of the near infrared identification of Carbon stars. I. The Local Group galaxies
 WLM, IC 10 and NGC 6822 \thanks{Based on observations obtained at the Italian Telescopio
Nazionale Galileo.}
}

%

   \author{P. Battinelli
          \inst{1}
\and
           S. Demers \inst{2}
\and
        F. Mannucci \inst{3}
         }


   \institute{
INAF, Osservatorio Astronomico di Roma
              Viale del Parco Mellini 84, I-00136 Roma, Italy\\
              \email {battinel@oarhp1.rm.astro.it }
         \and D\'epartement de Physique, Universit\'e de Montr\'eal,
                C.P.6128, Succursale Centre-Ville, Montr\'eal,
                                Qc, H3C 3J7, Canada\\
                \email {demers@astro.umontreal.ca }
	\and INAF - Instituto di Radioastronomia, Largo E. Fermi 5, 
	50125 Firenze, Italy\\
                \email{filippo@arcetri.astro.it}
}
   \date{Received; accepted}


   \abstract
{The selection of AGB C and M stars from NIR colours has been done in recent 
years using adjustable criteria that are in needs of standardization if one
wants to compare, in a coherent manner, properties of various populations.}
{We intend to assess the NIR colour technique to identify C and M stars.}
{We compare the NIR colours of several C stars previously identified 
from spectroscopy or narrow band techniques in WLM, IC 10 and NGC 6822.}
{We demonstrate that very few M stars have $(J-K)_0 > 1.4$ but a non 
negligible number of C stars are bluer than this limit. Thus, counts of
M and C stars based on such limit do not produce pure samples.}
{C/M ratios determined from NIR colours 
 must be regarded as underestimates
 mainly because the M 
numbers include many warm C stars and also K stars if no blue limit is
considered.}

\keywords{ galaxies, individual: IC 10, WLM, NGC 6822, Carbon stars
}

\titlerunning{WLM, IC 10 and NGC 6822}

\maketitle
%

\section{Introduction}

Carbon stars were initially identified in large numbers by objective prism surveys,
first in the Milky Way (Blanco 1965, Westerlund 1965) then a decade later similar
surveys toward the Magellanic Clouds (Blanco et al. 1978) yielded hundreds of
C stars. To reach fainter magnitudes thus larger distances, a photometric
technique based on two narrow band filters was introduced in the nineteen
eighties (Richer et al. 1984; Cook et al. 1986). This approach, based on the
($CN - TiO)$ index along with a colour, such as $(V-I)$ or $(R-I)$ has been
successfully exploited  to survey most of the Local Group galaxies (see for
example: Battinelli \& Demers 2005a; Brewer et al. 1995; Nowotny et al. 2003;
Rowe et al. 2005).
The narrow band technique presents, however, some serious drawbacks, the required 
filters are expensive to acquire and more importantly they are not
available on major telescopes. 

With the evolution of the near infrared (NIR)
instrumentation Asymptotic Giant Branch (AGB) stars became the subjects of a number of observations in
the Galaxy and beyond.
Mould \& Aaronson (1980) demonstrated that AGB stars in the SMC have
slightly bluer NIR colours that their cousins in the LMC. Spectroscopically
confirmed C stars were found to fall on an extend red tail in the $(J-K)$ vs $K$
plane, thus easily distinguishable from the O-rich M stars.
Survey of the literature reveals, however, that the C and M star border is 
ill-defined. Hughes \& Wood (1990) found from their NIR survey of  LMC Miras, with
known spectral types, that 98\% of O-rich stars have $J-K$ $<$ 1.6 while  15 out of 
87 C-rich stars have  $J-K$ $<$ 1.6. They therefore adopted a 
O- to C-rich
transition at $J-K=1.6$. 
 
 Davidge (2003) in his study of NGC 205 adopts, for C stars, 
 $(J-K)> 1.5 $ and $(H-K) > 0.4$ quoting
Hughes \& Wood (1990).
Analyzing the 2MASS data for the LMC, Nikolaev \& Weinberg  (2000)
set their blue limit of the C star region at $(J-K)$ $\approx$ 1.4.
More recently,  Cioni \& Habing (2003)  adopted  $(J-K)$ $>$ 1.4 and
 $(J-K)$ $>$ 1.3 for the LMC and SMC, respectively.
 However, Cioni \& Habing (2005) used another limit for 
the C stars in NGC 6822, namely $(J-K)_0$ $>$ 1.24, while Kang et al. (2006)
adopted for this galaxy $(J-K)_0$ $>$ 1.4 and $(H-K)_0 > 0.45$. 
For NGC 147  Sohn et al. (2006) took
 $(J-K)_0$ $>$ 1.25 and  $(H-K)_0 > 0.41$ for the C stars identification. 
The same team used  for NGC 185 the color limits
 $(J-K)_0$ $>$ 1.6 and $(H-K)_0 > 0.48$ (Kang et al. 2005). 
Davidge (2005) chose $(J-K)_0$ $>$ 1.4 and $(H-K)_0 > 0.45$ for both NGC 185 
and NGC 147.  Finally, Valcheva et al.
(2007) assumed $(J-K)_0$ $>$ 1.20  for the C stars in WLM.
Several of the cited authors set the color limit inspecting the $(J-K)_0$ 
color histogram.

It is well know that the NIR colours of the 
 RGB
 are function of the metallicity 
of the stellar population (Ferraro et al. 2000). 
The mean colours of O-rich or C-rich AGB stars brighter than the tip could 
similarly be metallicity dependent. There is at the present time no 
observational evidence for this effect. The NIR colour comparison of the AGB
in the Magellanic Clouds and a Galactic field (Schultheis et al. 2004) does not
reveal such trend.  
From this brief literature survey it appears evident that, 
 we are still 
far from any consensus about
the use of NIR photometry to select C and M AGB stars. 
The different colour limits adopted are often introduced to account to some extent
for metallicity 
differences of the parent galaxies.
 This could certainly explain why
the published C/M ratios for a given galaxy sometime  wildly differ.

In order to address this question we have started a program of 
JHK observations of several Local Group galaxies which already have
a known C star population obtained from the $(CN-TiO)$ index. The separation
between O-rich and C-rich AGB stars based on the $(CN-TiO)$ index has been
proved (Brewer et al. 1996, Albert et al., 2000) to be very reliable as 
long as the optical colors of the stars are redder than
a certain limit (e.g. $(R-I)_0 > 0.90$). For each galaxy, it is therefore 
reasonable to consider the sample of C stars identified with the narrow-band 
approach as a template to test other photometric criteria. 
In this first paper we discuss the case of three
galaxies of different metallicities: WLM and IC 10 with newly acquired data, and 
NGC 6822 already available in the literature.

\subsection{The target galaxies}

Wolf-Lundmark-Melotte (WLM) dwarf galaxy is located on the periphery 
of the Local Group and it is seen at a high Galactic latitude
($\ell = 76^\circ, b = -74^\circ$), thus being essentially extinction free. 
From the investigation of Dolphin (2000), we adopt $(m-M)_0$ = 24.90 for its
distance and a metallicity for its intermediate-age population of 
[Fe/H] = --1.4. WLM has been the target of a recent NIR study by Valcheva
et al. (2007) who identified numerous C and M AGB stars. They determined a 
C/M ratio for WLM that is quite different from the one calculated by
Battinelli \& Demers (2004) from the $(CN - TiO)$ criterion. 
As we shall see the difference comes mostly from the
different sets of  M and  C stars.

IC 10 is a dwarf irregular galaxy, most probably associated with M31 and
located at a rather low Galactic latitude ($\ell = 119^\circ, b = -3^\circ$).
It is often described as the only starburst galaxy of the Local Group. Its 
study is hindered by the high reddening $E(B-V) \approx$ 0.8 along the
line of sight. 
IC 10 is not extremely metal poor, from the oxygen abundance of its HII
regions Garnett (1990) determined [Fe/H] = --0.8, this value would
correspond to its youngest population but intermediate-age stars should
have slightly lower metallicities.
The central star forming region of IC 10 has been observed
in $JHK$ by Borissova et al. (2000) who, however, did not comment on the 
presence of C stars. Since the starburst makes the central region 
difficult to investigate, our NICS observations target an outer region.
IC 10 is particularly suited for NIR observations, it is relatively near 
at $(m-M)_0$ = 24.35, (Demers et al. 2004) and
 contains nearly 700 C stars distributed over an area much
larger than its starburst core. From Demers et al. (2004) we have the R,I
magnitudes of these C stars along with their narrow-band colors. 

For NGC 6822 we will use the NIR photometry published by Kang et al. (2006) and
the optical photometry by Letarte et. al. (2002).

\section{Colours of C stars}

When looking at spectra it is quite easy to distinguish a late M star from a 
C star. From the photometric point of view, the differences between the
two types are not so clear cut. We believe that the best non-spectroscopic 
way to divide M and
C stars is with the use of narrow band filters such as the $(CN-TiO)$ index.
A colour-colour diagram, based on this index and taken from Battinelli \&
Demers (2004) is shown in Figure 1. The upper branch, corresponding to C stars
is well isolated from the lower M star sequence. 
Investigation of the spatial distribution of the scattered points just above the M branch
reveals  that they corresponds to objects uniformly distributed over the CFH12K
field. We believe that they are non-stellar objects with sharpness just below the
rejection limit.
A very informative figure showing the CN and TiO wavelength ranges over
C and M spectra can be found in Nowotny et al. (2002).
This technique has limitations
however, it  fails
for bluer AGB stars, where the two branches converge into a big clump.
In this particular example there are few blueish C stars but this is not
always the case. For this reason we have adopted $(R-I)_0$ = 0.90 for the
blue limit of the C and M star counts. Furthermore, there is no easy way
to distinguish Galactic M dwarfs from giants without using time-consuming
multicolor systems (e.g. Majewski et al., 2000).

   \begin{figure}
   \centering
\includegraphics[width=7cm]{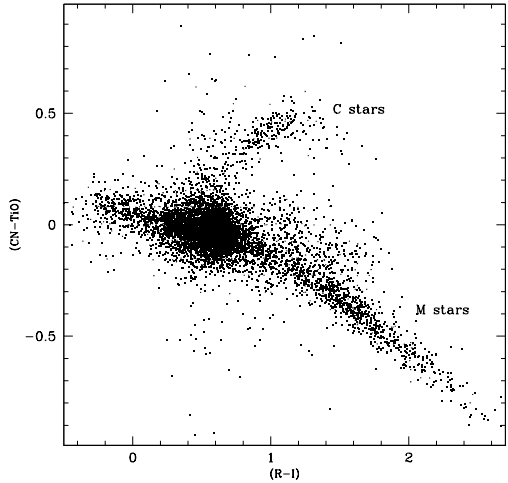}

   \caption{A typical colour-colour diagram 
(WLM, Battinelli \& Demers, 2004) showing the C and M branches.
}
              \label{cc diagram}
    \end{figure}

Figure 2, taken from Demers et al. (2002), displays a near infrared colour-magnitude 
diagram of the spectroscopically identified C stars in the Large Magellanic Cloud. 
Fig. 2 demonstrates that C stars, being distributed along the AGB, 
show an appreciable  $(J-K)$ colour range and  also
a large magnitude range. They are certainly not exclusively red. 
We also see that the use of C stars as standard candles requires
a severe restriction on the selected colour range. Indeed,
Weinberg \& Nikolaev  (2001), successfully used C stars selected in a {\it narrow colour range}
 as reliable standard candles to produce a 3D map of the LMC. 

The study of NIR colours of spectroscopically identified C stars is
unfortunately limited almost exclusively to the Magellanic Clouds. Indeed,
many Galactic C stars have NIR colours but because the reddening
in the plane is often far from  negligible their observed colours are not so useful. 
Furthermore, to select C stars, one needs an unbiased colour selection 
not always achieved. This is certainly the case for the Fornax dwarf
spheroidal galaxy where a few dozen C stars are known. Spectra of only the
very red giants were obtained to confirm their C star nature (Mould \& 
Aaronson 1980). 

   \begin{figure}
   \centering
\includegraphics[width=7cm]{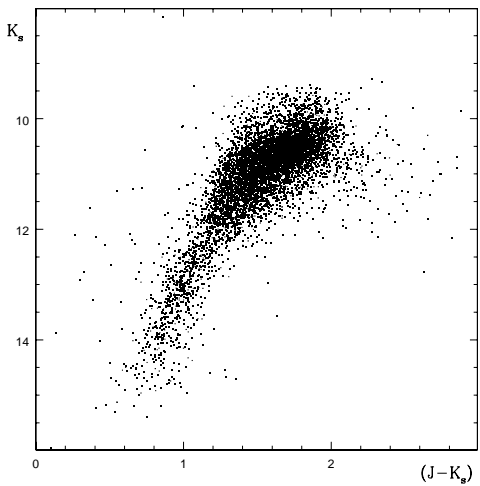}

   \caption{The CMD of spectroscopically identified C stars in the LMC
confirms their wide colour and magnitude ranges.
}
              \label{LMC CMD}
    \end{figure}

\section{Observations and reduction}

The $J,H,K_p$ observations were secured with NICS (Near Infrared Camera
Spectrometer; Baffa et al. 2001) installed at the Nasmyth focus of the TNG (Telescopio
Nazionale Galileo) on the island of La Palma. NICS is based on a HgCdTe
Hawaii 1024$\times$1024 array. The field of view for imaging is $4.2'
\times 4.2'$. 
One field in WLM $2.5'$ from its center and one $3'$ from the center of IC10 
were acquired
with a 4$\times$4 dithering pattern.
 The central core of each galaxy is therefore excluded.
 Table 1 summarizes
the observations obtained 
during two nights (IC 10: 2006-08-30; WLM: 2006-08-31)
under photometric conditions with pretty stable seeing $0.7''\div0.8''$.
   \begin{table}
      \caption[]{Journal of observations (J2000 coordinates)}
    $$
       \begin{array}{lccccc}
            \hline
            \noalign{\smallskip}
            {\rm Galaxy}&{\rm RA}&{\rm Dec}&{ J}&{ H}&{ K_s}  \\
           \noalign{\smallskip}
            \hline
            \noalign{\smallskip}

{\rm WLM}&00:02:02& -15:30:00&3840 s&1280 s&1280 s\\
{\rm IC 10}&00:19:55& +59:19:26&5760 s&1800 s&900 s\\
            \noalign{\smallskip}
            \hline
         \end{array}
     $$
   \end{table}
Data pre-reduction was performed with 
Speedy Near-infrared-data Automatic Pipeline (SNAP, Mannucci, in 
preparation),
described in http://www.tng.iac.es/news/2002/09/10/snap.
The basic steps of SNAP are briefly described here. After flat-fielding, a first-pass sky 
subtraction
is performed and the resulting images are combined together. Objects 
detected in this image are masked out  to perform a
second-pass sky subtraction, improving the estimate of the sky level and
the photometric accuracy. The final images are combined to a sub-pixel
precision by cross-correlating the object masks.

The photometric reduction of the combined images is done by fitting
model point-spread functions (PSFs) using DAOPHOT-II/ALLSTAR series of
programs (Stetson 1987, 1994). Since the resulting images are medians,
stars as bright as $K_s=13.5$ are not saturated. Furthermore, we see
no deviation from linearity when comparing the 2MASS magnitudes to the instrumental
magnitudes of the IC10 field.

\section{Results}

\subsection{WLM}

The line of sight toward WLM points far above the Galactic plane, thus our
small field of view contains few Galactic stars but  at faint
magnitudes one would expect to detect more unresolved galaxies than stars.
Cross identification with the 2MASS point source catalog yields only three
matches brighter than $K_s < 14.6$, nevertheless a reliable calibration of 
our instrumental magnitudes is possible. We obtain the following zero point
shifts:
$$ K_s = -0.442 (\pm 0.036) + K_{inst},$$
$$ J = 1.010 (\pm 0.060) + J_{inst},$$
$$ H = 0.311 (\pm 0.068) +H_{inst}.$$

DAOPHOT-II provides an image quality diagnostics SHARP.
Stetson (1987) devises a sharpness criterion by comparing the height of
the best-fitting Gaussian function to the height of a two-dimensional
delta function, defined by taking the observed intensity difference
between the central pixel of the presumed star image and the mean of the
remaining pixels used in the fit.
 For isolated
stars, SHARP should have a value close to zero, whereas for semi
resolved galaxies
and unrecognized blended doubles SHARP will be significantly greater than zero.
On the other end, bad pixels and cosmic rays produce SHARP less than zero.
SHARP
must be interpreted as a function of the apparent magnitude of all objects
because the SHARP parameter distribution 
{\bf 
degenerates
}
 near the magnitude limit;
see Stetson \& Harris (1998) for a discussion of this parameter. From Figure 3 
we define the stellar zone where SHARPs are within $\pm$ 0.15 from
zero. 
The color-magnitude diagram (CMD) of the field in WLM is displayed in 
Figure 4. Only stars with small SHARP parameters are plotted.
   \begin{figure}
   \centering
\includegraphics[width=6cm]{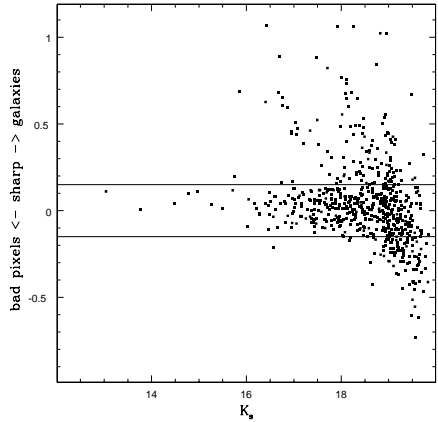}

   \caption{SHARP distribution of the objects identifies in the three bands
toward WLM. The two horizontal lines at $\pm 0.15$ limit the region of
stellar objects.
}
              \label{Sharp WLM}
    \end{figure}

   \begin{figure}
   \centering
\includegraphics[width=8cm]{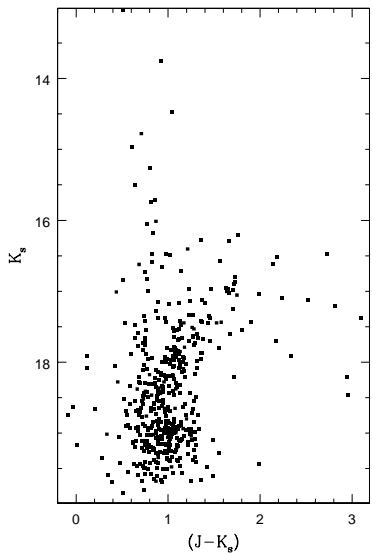}

   \caption{CMD of WLM field}

              \label{WLM}
    \end{figure}

Table 2 lists the near infrared magnitude and colors  of the 62 C stars 
previously identified.  
Magnitude errors are standard errors as defined by DAOPHOT  
thus not including the additional error from sky background variations.
We note however that the calibrations of our instrumental magnitudes with 2MASS
stars (particularly for IC 10 with 78 stars, next sub-section) yield calibration 
equations with slope 1 and dispersions along the trend of just a few hundreds 
of magnitudes 
thus suggesting a very efficient sky subtraction performed by SNAP.
 Colour errors are quadratic sums of the magnitude errors.
Optical photometry and 
identification number are from Demers \& Battinelli (2004).

   \begin{table*}
      \caption[]{Known C stars in the WLM NICS field{$^{\mathrm{a}}$}}
    $$
       \begin{array}{lccccccccc}
            \hline
            \noalign{\smallskip}
            {\rm id}&{ I}&{ (R-I)}&{(CN-TiO)}&{K_s}&{\sigma_K}&{(J-K_s)}&{\sigma_{JK}}&{(H-K_s)}&{\sigma_{HK}}  \\
           \noalign{\smallskip}
            \hline
            \noalign{\smallskip}

    23&    20.404&  1.233&  0.833& 17.428&  0.061&  1.904&  0.064&  0.731&  0.068\\
    24&    20.142&  1.532&  0.387& 16.795&  0.043&  1.730&  0.048&  0.609&  0.052\\
    26&    20.260&  1.026&  0.477& 17.729&  0.076&  1.293&  0.077&  0.362&  0.080\\
    29&    20.230&  1.083&  0.486& 17.428&  0.063&  1.364&  0.065&  0.347&  0.066\\
    31&    20.145&  0.965&  0.405& 17.885&  0.104&  1.114&  0.108&  0.154&  0.112\\
    38&    20.238&  0.971&  0.363& 17.665&  0.086&  1.147&  0.088&  0.239&  0.090\\
    41&    19.956&  1.218&  0.458& 17.446&  0.073&  1.526&  0.077&  0.476&  0.083\\
    43&    19.995&  0.931&  0.451& 17.531&  0.063&  1.281&  0.065&  0.285&  0.068\\
    44&    20.373&  1.000&  0.454& 17.937&  0.087&  1.178&  0.090&  0.268&  0.092\\
    45&    21.104&  1.128&  0.446& 17.255&  0.051&  1.703&  0.053&  0.549&  0.055\\

            \noalign{\smallskip}
            \hline
         \end{array}
     $$
\begin{list}{}{}
\item[$^{\mathrm{a}}$] Table 2 is presented in its entirety in the electronic 
edition of Astronomy \& Astrophysics.
A portion is shown here for guidance regarding its
form and content.
\end{list}
   \end{table*}

\subsection{IC 10}

Thanks to its low Galactic latitude, the IC 10  NICS field contains 78 2MASS stars 
suitable for magnitude calibrations.
We obtain the following zero points:

$$K_s = -0.264 (\pm 0.032) + K_{inst},$$
$$ J = 1.020 (\pm 0.011) +J_{inst},$$
$$H = 0.299 (\pm 0.022) +H_{inst}.$$

The CMD of the region observed in IC 10 is displayed in Figure 5.
We find in the NICS field 52 C stars, listed in Table 3,  previously identified by Demers et 
al. (2004). According to our reddening map, the mean reddening in the
NICS field is $E(B-V) = 0.86$ which corresponds following the relations of
Schlegel et al. (1998) to $E(J-K) = 0.45$, $E(H-K) = 0.23$ and A$_K$ = 0.32.
The optical magnitudes and colours of IC 10 are individually deredden
using our reddening map.

   \begin{figure}
   \centering
\includegraphics[width=7cm]{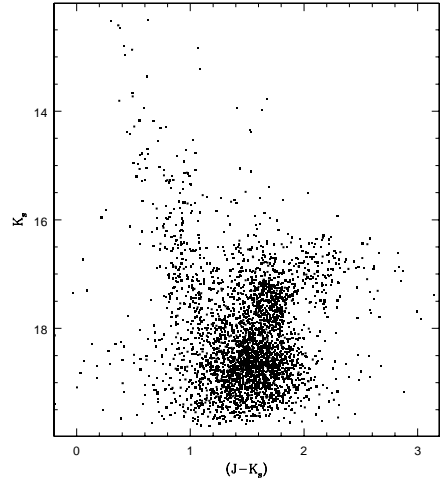}

   \caption{CMD of IC 10 field}

              \label{ic 10}
    \end{figure}

   \begin{table*}
      \caption[]{Known C stars in the IC 10 NICS field{$^{\mathrm{a}}$}}
    $$
       \begin{array}{lccccccccc}
            \hline
            \noalign{\smallskip}
            {\rm id}&{ I_0}&{(R-I)_0}&{(CN-TiO)}&{ K_s}&{\sigma_K}&{(J-K_s)}&{\sigma_{JK}}&{(H-K_s)}&{\sigma_{HK}}  \\
           \noalign{\smallskip}
            \hline
            \noalign{\smallskip}
    51&     20.085&     0.911&     0.531&    17.036&     0.051&     2.220&     0.055&     0.831&     0.072\\
    58&     19.306&     1.193&     0.373&    16.494&     0.032&     2.089&     0.037&     0.713&     0.038\\
    60&     19.571&     1.194&     0.512&    16.720&     0.038&     2.067&     0.043&     0.752&     0.045\\
    62&     20.092&     0.985&     0.518&    17.247&     0.051&     1.989&     0.055&     0.694&     0.062\\
    63&     20.186&     1.037&     0.530&    17.147&     0.047&     2.247&     0.053&     0.762&     0.058\\
    68&     19.651&     0.900&     0.410&    17.237&     0.056&     1.732&     0.061&     0.552&     0.067\\
    70&     20.183&     1.203&     0.638&    16.879&     0.041&     2.533&     0.046&     1.052&     0.051\\
    71&     19.253&     0.992&     0.687&    16.606&     0.032&     2.039&     0.036&     0.717&     0.039\\
    73&     20.213&     1.088&     0.480&    16.900&     0.039&     2.234&     0.043&     0.853&     0.049\\
    75&     19.662&     0.963&     0.671&    17.362&     0.062&     1.770&     0.065&     0.524&     0.068\\

            \noalign{\smallskip}
            \hline
         \end{array}
     $$
\begin{list}{}{}
\item[$^{\mathrm{a}}$] Table 3 is presented in its entirety in the electronic 
edition of Astronomy \& Astrophysics.
A portion is shown here for guidance regarding its
form and content.
\end{list}
   \end{table*}

   \begin{figure*}
   \centering
\includegraphics[width=9cm]{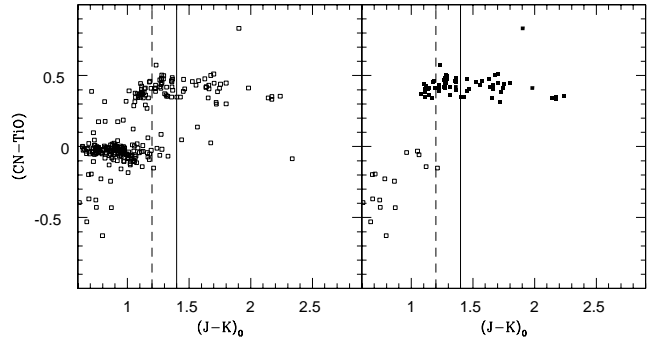}

   \caption{Left panel: color-color plot of the WLM AGB stars seen in our NICS field. The two vertical lines
correspond to the 1.2 and 1.4 $(J-K_s)_0$ limits. Right panel: C (filled squares) and M
 stars as defined from the RICNTiO photometry. In both panels only stars with K magnitudes
brighter than the TRGB are plotted.}

              \label{WLMjkcn}
    \end{figure*}
  \section{Discussion}
To compare the NIR properties of C and M stars identified by the narrow-band 
technique, we show in the left panel of Figure 6  an aspect of the 
cross-identification 
between our CFHT photometry of WLM (all stars) and our new NIR observations. 
We see two well separated groups of points: the C stars with $(CN-TiO)>0.3$ 
and the K and M giants with negative $(CN-TiO)$. 
The solid vertical line at $(J-K_s)_0=1.4$ 
is the often adopted limit for the C star selection (see Sect. 1). 
Stars with $(CN-TiO)$ smaller than $\approx -0.3$ have $(J-K_s)_0$ and $(H-K_s)_0$ 
colors typical of late dwarfs (see Bessell \& Brett, 1988) 
that are seen here along the line of sight and mimic AGB stars.
The dashed line is the Valcheva et al. (2007) limit to select C stars from the AGB stars 
found in their 
WLM study. They called AGB stars all the objects brighter than the TRGB. 
 We must stress, however, that Valcheva et al. 
(2007) consider  all the AGB stars bluer than $(J-K_s)_0=1.2$  as M stars. 
We see from Fig. 6 that their M star sample is certainly polluted by a 
number of C stars and certainly by stars earlier than spectral type M0.
The right panel of Fig 6. includes only C and M stars as defined from the $RICNTiO$ 
photometry (Battinelli \& Demers, 2004), i.e: C stars with 
$(CN-TiO)>0.3$ and $(R-I)_0>0.9$ and M stars with $(CN-TiO)<0$ and 
$(R-I)_0>0.9$.  From the comparison 
of the two panels two facts emerge: 1)  while the $(J-K_s)_0$ color threshold is 
appropriate as red limit for M stars it is not suitable for the blue limit of
C stars. This limit implies that some C stars are misidentified as M-type; 
2) the adoption of the $(R-I)_0=0.9$ threshold does not show a similar 
drawback, we see that  nearly all the stars in the upper group are considered 
as C stars.  On the other hand this limit cuts drastically the number 
of stars in the lower group. This is not surprising since stars with 
$(J-K)_0 < 1.0$ have spectral types earlier than M0 (Bessell \& Brett 1988).

   \begin{figure*}
   \centering
\includegraphics[width=9cm]{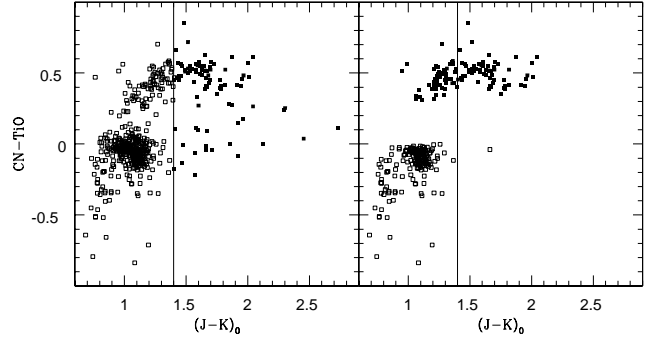}

   \caption{Left panel: color-color plot of the M and C stars (filled squares) 
as defined by Kang et al. (2006) for NGC 6822.  
The  vertical line
correspond to the  $(J-K_s)_0=1.4$ limit. Right panel: C (filled squares) 
and M stars as defined from the RICNTiO photometry.}

              \label{n6822jkcn}
    \end{figure*}
A similar plot, shown in Figure 7,  has been obtained for NGC 6822 by 
matching the published lists of C and M stars identified from NIR 
photometry by of Kang et al. (2006) with Letarte et al. (2002) database. 
We see that for the stars in the upper group this galaxy
behaves similarly to WLM.   On the other hand, the adoption of the $(R-I)_0=0.9$ 
threshold does not cut seriously the number of stars in the bottom group, 
contrary to the case of WLM. 
This difference might 
very well be due to the  higher metallicity of NGC 6822 that makes the 
RGB and AGB redder. 
This figure suggests $(J-K_s)_0=1.2$ as an appropriate limit for the C-M
separation contrary to the $(J-K_s)_0=1.4$ adopted by Kang et al. (2006) on the 
basis of the color histogram.
We notice the presence in the left panel of a number of red stars with 
$(CN-TiO)\approx 0$ that disappear  in the right panel. These objects 
were therefore matched to stars with  $(R-I)_0<0.9$. The easiest explanation
for them is possible mismatches (within 1 arcsec).
However a similar population is also visible in Fig. 1. From the spatial
distribution of these odd objects we conclude they are background galaxies
with sharpness small enough to mimic real stars.

   \begin{figure*}
   \centering
\includegraphics[width=9cm]{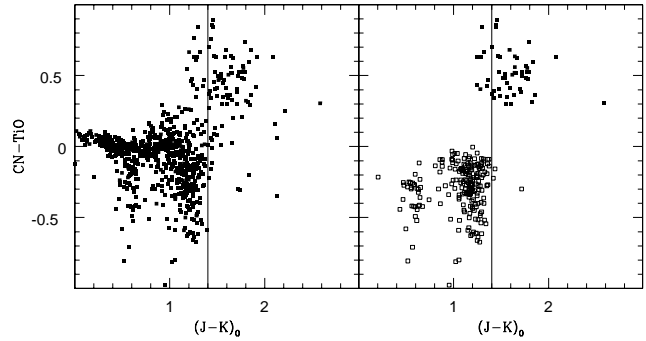}

   \caption{Same as Fig. 6 for IC 10. }

              \label{ic10jkcn}
    \end{figure*}

Contrary to the two previous galaxies, nearly all C stars identified by
 the narrow band technique in IC 10 (see Figure 8) 
have  $(J-K_s)_0>1.4$. We see in this galaxy the presence of several objects with 
very negative $(CN-TiO)$. Such extreme negative $(CN-TiO)$ were already 
found in M 31 by Battinelli et al. (2003). 

There is no doubt that numerous foreground red stars contribute to the
large number of M stars seen in Fig. 8. As Demers et al. (2004) have shown,
an accurate estimate of this foreground would be needed to properly
determine the C/M ratio. We, however, do not need to know here this
contribution since our aim is not to use our NIR to determine the C/M ratios
but to assess the NIR approach.

We find in the literature three other Local Group galaxies which C star 
populations have been identified using NIR colors, namely NGC 147 
(Sohn et al., 2006), NGC 185 (Kang et al. 2005) and NGC 205 (Davidge 2003).
Unfortunately, the full lists of AGB stars in these galaxies are
 not available since the authors  published no list or only the list of 
the identified C stars.
There are no $(CN-TiO)$ observations of the Magellanic Cloud AGB stars.

We conclude, from what has been discussed above, that a color $(J-K_s)_0=1.4$ 
can be regarded as a conservative  limit for the selection of C stars among 
the AGB stars. Indeed, stars selected 
according to this criterion consist exclusively of 
C stars even though a non negligible number
of genuine C stars, bluer than this adopted limit, may not be counted (depending 
on the metallicity). 
Another drawback of NIR colors compared to the narrow-band approach is the 
selection of AGB M stars. 
Beside the fact that a number of C stars are misidentified as M-type, very 
often all the stars above the TRGB and bluer than the above limit are 
considered as AGB M stars while obviously a blue $(J-K_s)_0$ color limit 
should be introduced to omit late K stars. The determination
of such blue threshold is far from straightforward lacking a tight relation 
between $(J-K_s)_0$ colors and spectral types. The neglect of a blue 
limit to select AGB M stars led Valcheva et al. (2007) to largely
overestimate the number of M stars. For instance, Valcheva et al. (2007) counted as 
M stars objects as blue as $(J-K_s)_0=0.5$ while such colour, according to Bessell \& 
Brett (1988), corresponds to G2  giants. This obviously explains why their C/M 
ratio is much smaller than the value obtained by
Battinelli \& Demers (2004) using narrow band photometry. 
For all these reasons, C/M ratios deduced from the use of NIR colors 
significantly underestimate their real values.

Similar conclusions can be reached by comparing the $(H-K_s)_0$ colors of 
C and M stars. 
We found a $(H-K_s)_0\approx 0.4$ limit is the counterpart of the $(J-K_s)_0>1.4$ 
even though not as clearly defined as the latter. 

   \begin{table*}
      \caption[]{Mean NIR properties of C stars with $<(J-K_s)>_0 > 1.4$}
    $$
       \begin{array}{llccccccc}
            \hline
            \noalign{\smallskip}
            {\rm Galaxy}&{\rm [Fe/H]}&{\rm N_C}&{\rm \langle K_{s,0}\rangle}&{\rm \langle (J-K_s)\rangle_0}&{\rm \langle (H-K_s)\rangle_0}&{\rm \langle M_{K_s}\rangle}&{\rm \langle R-J\rangle}&{\rm T_{eff}}  \\
           \noalign{\smallskip}
            \hline
            \noalign{\smallskip}
{\rm LMC}&-0.5&4617&10.59&1.65&0.56&-8.01&&\\
{\rm IC 10}&-0.8& 212&16.77&1.74&0.57&-7.68&2.42&3400\\
{\rm NGC 147}&-1.0& 77&16.80&1.90&0.79&-7.60&2.67&3200\\
{\rm Fornax}&-1.0&26&13.07&1.61&0.57&-7.67&&\\
{\rm SMC}&-1.1&317&11.18&1.62&0.58&-7.92&&\\
{\rm NGC 6822}&-1.25&141& 15.85&1.77&0.75&-7.51&2.61&3250\\
{\rm NGC 185}&-1.3& 73&16.19&2.25&0.86&-7.93&0.17&?\\
{\rm WLM}&-1.4& 38&17.19&1.89&0.69&-7.71&2.39&3400\\
            \noalign{\smallskip}
            \hline
         \end{array}
     $$
   \end{table*}

\subsection{Mean NIR properties of C stars}

In Table 4 we list the average NIR properties of the C stars identified in 
each galaxy by adopting the $(J-K_s)_0>1.4$ criterion for stars brighter than 
the TRGB. Distance moduli and  [Fe/H] are from Battinelli \& Demers (2005a).
 For WLM and IC 10, we determine the 
magnitude of the tip using the adopted [Fe/H] and the calibration 
published by Ivanov \& Borissova (2002). Data for the LMC, SMC and Fornax
are from Demers et al. (2002) while those for the other galaxies of from
references cited above.
The NGC 185 colours stand out as being quite red. A NIR survey of its
AGB was done by Davidge (2005) who quotes a colour difference with
Kang et al. (2005) of $\Delta(J-K_s)$ = 0.21. We cannot cross identify
Davidge's C stars with ours because their coordinates are not available.

In the last column we give the mean effective temperature calculated
from Loidl et al.  (2001) from the $\langle R - J\rangle$ of C stars. 
It is not too surprising that the C stars in each galaxy are very similar 
temperature wise. Only the redder stars, with $(J-K_s)_0>1.4$ are selected
via the NIR approach. We note that the J magnitudes of C stars in NGC 185
as published by Kang et al. (2005) are too large making the $(R-J)$ so
small that no temperature can be calculated. 

\section{Conclusions}

We have shown that the samples of C and M stars selected from NIR
colors differ significantly from those obtained using the $RICNTiO$
photometry. The main differences are: 
\par {\it i)}
The NIR sample of M stars is polluted by a significant number of C
stars misidentified as M. This is not the case for $RICNTiO$ identified C
stars where the $(CN-TiO)$ color
is an effective discriminant. 
\par {\it ii)} 
Both NIR and $RICNTiO$ selection criteria for M stars require the
adoption
of a blue threshold to weed out K stars from the sample. In the $RICNTiO$ approach,
a limit of
$(R-I)_0=0.9$ is generally adopted. Not all the authors using NIR color
consider a similar blue limit, thus ending up with a large overestimate
of the M star number.  

Similarly, Groenewegen (2004), from a study of spectroscopically
classified long period variables in the Magellanic Clouds, concludes that
$(J-K)=1.4$ cannot be use to properly separate M and C variables.
 From the above considerations it is evident that the C/M ratios
obtained
from NIR and narrow band photometry can be very different. It is
therefore
not justified to adopt C/M vs [Fe/H] calibrations obtained from RICNTiO
to convert
NIR C/M into metallicities. A correct approach would be to calibrate the
NIR C/M
in terms of [Fe/H] similarly to what Battinelli \& Demers (2005a) did
for
narrow band C/M.  

 On the basis of the data used in this paper the mean properties of C stars
identified with NIR colors do not seem to be significantly sensitive to
the metallicity of the parent galaxy. In particular, contrary to the
 $\langle M_I \rangle$
of C stars that has been proved to be fairly constant (Battinelli \&
Demers, 2005b),  the  $\langle M_{K_s} \rangle$ shows a wide range of variation.

\begin{acknowledgements}
This research
is funded in parts (S. D.) by the Natural Sciences and Engineering Research
Council of Canada. 

This publication makes use of data products from the Two Micron All Sky Survey,
which is a joint project of the University of Massachusetts and the Infrared
Processing and Analysis Center/California Institute of Technology, funded by
the National Aeronautics and Space Administration and the National Science Foundation.
\end{acknowledgements}


\begin{thebibliography}{}


\bibitem[2000] {alb00} Albert, L., Demers, S., \& Kunkel, W. E., 2000,
AJ, 119, 2780

\bibitem[2001] {baf01} Baffa, C., Comoretto, G., Gennari, S. et al. 2001,
A\&A, 378, 722

\bibitem[2003] {bat03} Battinelli, P., Demers, S. \& Letarte, B., 2003, AJ, 125, 1298

\bibitem[2004] {bat04} Battinelli, P. \& Demers, S. 2004, A\&A, 416, 111

\bibitem[2005] {bat05} Battinelli, P. \& Demers, S. 2005a, A\&A, 434, 657

\bibitem[2005] {bat05b} Battinelli, P. \& Demers, S. 2005b, A\&A, 442, 159

\bibitem[1988] {bes88} Bessell, M. S.   \& Brett, J. M., 1988, PASP, 100, 1134

\bibitem [1965] {bla65} Blanco, V. M. 1965 in {\it Stars and Stellar Systems}
vol. V, eds. A. Blaauw and M. Schmidt, University of Chicago Press,
Chicago, p241.

\bibitem [1978] {bla78} Blanco, B. M., Blanco, V. M., \& McCarthy, M. F.
1978, Nature, 271, 638

\bibitem[2000] {bor00} Borissova, J. Georgiev, L., Rosado, M. et al. 2000,
A\&A, 363, 130

\bibitem [1995] {bre95} Brewer, J. P., Richer, H. B., \& Crabtree, D. R. 1995,
AJ, 109, 2480

\bibitem [1996] {bre96} Brewer, J. P., Richer, H. B., \& Crabtree, D. R. 1996,
AJ, 112, 491


\bibitem[2003]{cio03} Cioni, M.-R. L., \& Habing, H. J. 2003, A\&A, 402, 133

\bibitem[2005]{cio05} Cioni, M.-R. L., \& Habing, H. J. 2005, A\&A. 429, 837

\bibitem[1986] {coo86} Cook, K. H., Aaronson, M., \& Norris, J. 1986,
ApJ, 305, 634


\bibitem[2003] {dav03} Davidge, T. J. 2003, ApJ, 597, 289

\bibitem[2005] {dav05} Davidge, T. J. 2005, AJ, 130, 2087


\bibitem[2002] {dem02} Demers, S., Dallaire, M., \& Battinelli, P. 2002, AJ.
123, 3428

\bibitem[2004] {dem04}  Demers, S., Battinelli, P., \& Letarte, B. 2004, A\&A,
424, 125
\bibitem[2000] {dol00} Dolphin, A. E. 2000, ApJ, 531, 804

\bibitem[2000]{fer00} Ferraro, F. R., Montegriffo, P. A., \& Fusi-Pecci, F.
2000, AJ, 119, 1282

\bibitem[1990] {gar90} Garnett, D. R. 1990, ApJ, 363, 142

\bibitem[2004]{gro04} Groenewegen, M. A. T. 2004, A\&A, 425, 595

\bibitem[1990] {hug90} Hughes, S. M. G., \& Wood, P. R. 1990, AJ, 99, 784

\bibitem[2002] {iva02} Ivanov, V. D., \& Borissova, J. 2002, A\&A, 390, 937

\bibitem[2005] {kan05} Kang, A., Sohn, Y.-J., Rhee, J., et al. 2005, A\&A, 437,
61

\bibitem[2006] {kan06} Kang, A., Sohn, Y.-J., Kim, H.-L. et al. 2006, A\&A,
454,717

\bibitem[2002] {let02} Letarte, B., Demers, S., Battinelli, P., \&
 Kunkel, W. E. 2002, AJ, 123, 832

\bibitem[2001] {loi02} Loidl, R., Lan\c con, A., \& J\o rgensen, U. G. 2001, A\&A, 371, 1065

\bibitem[2000]{maj00} Majewski, S. R., Ostheimer, J. C., Patterson, R. J., et al. 2000, AJ, 119, 760

\bibitem[1980]{mou80} Mould, J., \& Aaronson, M. 1980, ApJ, 240, 464

\bibitem[2000]{nik00} Nikolaev, S., \& Weinberg, M. D. 2000, ApJ, 542, 804

\bibitem [2002] {now02} Nowotny, W., \& Kirschbaum, F. 2002, Hvar Obs.
Bull. 26, No. 1, p63

\bibitem[2003]{now03} Nowotny, A., Kerschbaum, F., Olofsson, H., \&
Schwarz, H. E. 2003, A\&A, 403, 93

\bibitem[2005] {row05} Rowe, J., Richer, H., Brewer, J., \& Crabtree, D.
2005, AJ, 129, 729

\bibitem[1984] {ric84} Richer, H. B., Crabtree D. R., \& Pritchet, C. J.
1984, ApJ, 287, 138

\bibitem[1998]{sch98} Schlegel, D., Finkbeiner, D., \& Davis, M.
1998, ApJ, 500, 525

\bibitem[2004]{shc04} Schultheis, M., Glass, I. S., \&, Cioni, M.-R., 2004,
A\&A, 427, 945

\bibitem[2006] {soh06} Sohn, Y.-J., Kang, A., Rhee, J. et al. 2006, A\&A, 445,
69

\bibitem[1987]{ste87}Stetson, P. B. 1987, PASP, 99, 191

\bibitem[1988] {ste88} Stetson, P. B. \& Harris, W. E. 1988, AJ, 96, 909

\bibitem[ 1994]{ste94} Stetson, P. B. 1994, PASP, 106, 250

\bibitem[2007] {val07} Valcheva, A. T., Ivanov, V. D., Ovcharov, E. P., \&
            Nedialkov, P. L. 2007, A\&A, 466, 501 
\bibitem[2000]{nik00} Weinberg, M. D.,  \& Nikolaev, S., 2001, ApJ, 548, 712

\bibitem[1965]{wes65} Westerlund, B. E. 1965, MNRAS, 130, 45


\end{thebibliography}
\end{document}